\documentclass[12pt,twoside]{article}
\usepackage{amsmath}
 \usepackage{amssymb} 
\usepackage{amsmath,amscd,graphics}

 \def\be#1\ee{\begin{equation}#1\end{equation}}
 \def\bay#1\eay{\!\!\!\left\{\!\!\begin{array}{l}#1\displaystyle\end{array}\right.}
 \def\bln#1\eln{\begin{array}{l}#1\displaystyle\end{array}}

 \def\bma#1\ema{{\allowdisplaybreaks\begin{align}#1\end{align}}}
 \def\nnm{\notag}

 \def\bgr#1\egr{{\allowdisplaybreaks\begin{gather}#1\end{gather}}}
 \def\ef#1{(\ref{#1})}
 \def\qef#1{$(\ref{#1})$}

       \newtheorem{lemma}{\bf Lemma}[section]
       \newtheorem{theorem}[lemma]{\bf Theorem}

       \newtheorem{remark}[lemma]{\bf Remark}

\begin{document}

\title{{\LARGE
\textbf{Algebraic time-decay for the bipolar quantum hydrodynamic
model\thanks{Corresponding author: Hai-Liang Li}}}
}
\author{\textbf{Hai-Liang Li$^{1)}$,~~Guojing Zhang$^{2)}$,
~~Kaijun Zhang$^{2)}$}\\[2mm]
{\it\small $^{1)}$Department of Mathematics, Capital Normal
University~~~~~~~}\\
{\it\small Beijing 100080, P. R. China }\\
{\it\small $^{2)}$School
of Mathematics and Statistics,
 Northeast Normal University}\\
  {\it\small Changchun 130024, P.R.China }\\
{\it\small  email:~hailiang.li.math@gmail.com $($H.L$)$,
zhanggj112@nenu.edu.cn $($G.Z$)$}
\\
{\it\small  zhangkj201@nenu.edu.cn $($K.Z$)$}}\date{}
 \maketitle
\vspace{-0.5cm}
\renewcommand{\thefootnote}{\fnsymbol{footnote}}
 \pagestyle{myheadings}
 \markboth{Algebraic time-decay for the bipolar QHD}
 {H.-L. Li, G.-J. Zhang, $\&$ K.-J. Zhang}

 \begin{abstract}
\normalsize{\noindent 
The initial value problem is considered in the present paper for
bipolar quantum hydrodynamic model for semiconductors (QHD) in
$\mathbb{R}^3$. We prove that the unique strong solution exists
globally in time and tends to the asymptotical state with an
algebraic rate as $t\rightarrow+\infty$. And, we show that the
global solution of linearized bipolar QHD system decays in time at
an algebraic decay rate from both above and below. This means in
general, we can not get exponential time-decay rate for bipolar
QHD system, which is different from the case of unipolar QHD model
(where global solutions tend to the equilibrium state at an
exponential time-decay rate) and is mainly caused by the nonlinear
coupling and cancelation between two carriers. Moreover, it is
also shown that the nonlinear dispersion does not affect the long
time asymptotic behavior, which by product gives rise to the
algebraic time-decay rate of the solution of the bipolar
hydrodynamical model in the semiclassical limit.}
\end{abstract}

\textbf{Key words}: \begin{minipage}[t]{120mm} Quantum
hydrodynamics; Algebraic decay rate.
\end{minipage}
\bigskip

\section{Introduction}
The quantum hydrodynamic(QHD) model for semiconductors is derived
and studied recently in the modelings and simulations of
semiconductor devices, where the effects of quantum mechanics
arises. \textbf{The basic observation concerning the quantum
hydrodynamics is that the energy density consists of one additional
new quantum correction term of the order $O(\varepsilon)$ introduced
first by Wigner~\cite{Wigner32} in 1932, and that the stress tensor
contains also an additional quantum correction
part~\cite{AncTie87,AncIaf89} related to the quantum Bohm potential
(or internal self-potential)~\cite{Bohm52} \be Q(\rho) =
-\frac{\varepsilon^2}{2m}\frac{\Delta\sqrt{\rho}}{\sqrt{\rho}},\label{disp-2}
\ee with observable $\rho>0$ the density, $m$ the mass, and
$\varepsilon$ the Planck constant.} The quantum potential $Q$ is
responsible for producing the quantum behavior. Such possible
relation was also implied in the original idea initialized by
Madelung~\cite{Madelung27} in 1927 to derive quantum fluid-type
equations in terms of Madelung's transformation applied to wave
function of Schr\"odinger equation of pure state. Recently, the
moment method is employed to derive quantum hydrodynamic equations
for semiconductor device at nano-size based on the Wigner-Boltzmann
(or quantum Liouville) equation, see in \cite{MRS1990} for details.
For derivation about quantum hydrodynamical equations and related
quantum models, one can refer to
\cite{DR2003,Gardner1994,Jungel2001} and the reference therein.
\par

In this paper, we consider the Cauchy problem of the bipolar quantum
hydrodynamic(QHD) model for semiconductors in
$\mathbb{R}^3\times[0,+\infty)$ which reads \bgr
\partial _t\rho_i+\nabla \cdot
(\rho_iu_i)=0,\label{1.1}\\
 \partial_t(\rho_i u_i)+\nabla\cdot(\rho_iu_i \otimes u_i)+
\nabla P_i(\rho_i)=q_i\rho_iE+\frac{\varepsilon
^2}{2}\rho_i\nabla(\frac{\triangle\sqrt{\rho_i}}{\sqrt{\rho_i}})-\frac{\rho_iu_i}{\tau_i},\label{1.2}\\
\lambda^2\nabla\cdot E=\rho_a-\rho_b-\mathcal{C}(x),~\nabla \times E
=0,~E(x)\rightarrow 0,~|x|\rightarrow +\infty,\label{1.3} \egr with
the initial conditions \be(\rho_i
,u_i)(x,0)=(\rho_{i_0},u_{i_0})(x),\label{1.4} \ee where the index
$i=a,b$ and $q_a=1,~q_b=-1$. The variables $\rho_a>0,\rho_b>0$ and
$u_a,u_b$ and $E$ are the particle densities, velocities and
electric field, respectively. We can define the usual momentum
$J_a,J_b$ as $ J_a=\rho_au_a,\  J_b=\rho_bu_b$. $P_a(\cdot)$ and
$P_b(\cdot)$ are the pressure-density functions. The parameters
$\varepsilon>0$, $\tau_a=\tau_b=\tau>0$, and $\lambda>0$ are the
scaled Planck constant, momentum relaxation time, and Debye length
respectively. $\mathcal{C}=\mathcal{C}(x)$ is the doping profile
function. When it holds $(\rho_b,\rho_bu_b)\equiv(0,0)$ formally,
the above model reduces the unipolar quantum hydrodynamical model.
\par Recently, many mathematical efforts are made on the study of
the QHD model for semiconductors on both the steady state solutions
and the evolutional (time-dependent) solutions. The investigation on
unipolar QHD model are well-understood up to now. The steady state
solutions of unipolar QHD model are studied in
\cite{{bgms},{GJ773-800},{J463-479},{zJ845-856}} in one-dimensional
or multi-dimensional bounded domain for different boundary
conditions, and the steady state solution of the unipolar viscous
quantum hydrodynamical system is investigated in \cite{GJ183-203}.
For the one-dimensional time-dependent case, the short time
existence of solutions of unipolar model \cite{HJ20031-15} and the
global existence theory with the exponential stability of stationary
state in whole space \cite{{JL2003},{JL2004},{HLM2003}} are
established. For the multi-dimensional case,  the local existence of
solutions is obtained for irrotational fluid \cite{JM485-495}, and
the local and global existence theory and  exponential stability of
equilibrium state analysis are also investigated for irrotational
fluid on spatial periodic domain \cite{LM215-247}. The corresponding
existence theory for time-dependent solution for general rotational
fluid is usually difficult and is obtained very recently in
\cite{HLMO2003}, where the exponential decay to the stationary state
obtained therein is made. Moreover, the asymptotical small scaling
analysis including the relaxation time limit, small Debye length
limit and the semiclassical limit for the global solutions are
studied in \cite{{JL2005},{LL195-212},{ZLZ2006}} respectively.
\par

However, the results for bipolar QHD model are quite fewer compared
with those obtained for unipolar QHD model. So far, only the  steady
state solutions are studied partially in
\cite{{Lz2005},{ZZ2005},{U69-88}} for bounded and unbounded domain,
and the semiclassical limit and relaxation limit of the
global-in-time solutions are investigated in \cite{ZLZ2006}, where
the global existence of time-dependent solution is also proven, but
without the deriving the large time behavior. The main difficulty in
dealing with the bipolar QHD model is the coupling and interaction
between the two carriers, which may cause some cancelation, and it
is not clear that the equilibrium state to the bipolar QHD is still
exponential stable or not for small perturbation.
 \par

 In this paper, we study
the time-decay rate of global solutions to the Cauchy problem for
the bipolar QHD \ef{1.1}--\ef{1.4} in $\mathbb{R}^3$. We shall show
that the solution to the IVP for bipolar QHD tends to the
equilibrium state at an algebraic decay rate. This property is
different from the unipolar QHD model and is caused by the
interaction and nonlinear coupling of the two carriers which make
the convergence of solution to the equilibrium state slower.
\par

We have the following main result.

\begin{theorem}{\label{theorem 1.1}}
Assume $\mathcal{C}(x)= c^*$ with $c^*$ a positive constant, and
$\rho_a^*>0,~\rho_b^*>0$ are constants satisfying
$\rho_a^\ast-\rho_b^\ast-c^*=0$. Assume $P_a, \ P_b\in C^6$ and $
~P_a^\prime(\rho_a^\ast),~P_b^\prime(\rho_b^\ast)>0$.
Let the initial data satisfy $(\rho_{i_0}-\rho_i^*,u_{i_0})
 \in {H^6(\mathbb{R}^3)}\times{\mathcal{H}^5(\mathbb{R}^3)}$, $i=a,b,$
  with
$\Lambda_0:=\|(\rho_{i_0}-\rho_i^\ast, u_{i_0})
 \|_{H^6(\mathbb{R}^3)\times{\mathcal{H}^5(\mathbb{R}^3)}}.
 $
\textbf{Then, there exists $\Lambda_1>0$ such that if
$\Lambda_0\leq\Lambda_1$, the unique solution $(\rho_i,u_i,E)$ of
the IVP~$(\ref{1.1})$-\qef{1.4} with $\rho_i>0$ exists globally in
time and satisfies for $i=a,b$ that \be
 ( \rho_i - \rho_i^\ast )\in C^k(0,T;H^{6-2k}(\mathbb{R}^3)),\
 u_i\in C^k(0,T;\mathcal{H}^{5-2k}(\mathbb{R}^3)),\nnm
 \ee
 \be
 E\in C^k(0,T;\mathcal{H}^{6-2k}(\mathbb{R}^3)),
 \label{1.7}
 \ee
 for $k=0,1,2$.\\
Moreover, the solution $(\rho_i,u_i,E)$ tends to the equilibrium
state $(\rho_i^\ast,0,0)$ at an algebraic time-decay rate
 \bgr
  (1+t)^k\| D^k(\rho_i - \rho_i^\ast)\|^2
 +(1+t)^5\| \varepsilon D^6(\rho_i - \rho_i^\ast)\|^2
 \leq
 c\Lambda_0,\ \ 0\leq k\leq5,
\label{1.8}
\\
 (1+t)^k\| D^k (u_i,J_i)\|^2+(1+t)^k\|
 D^k E\|^2+(1+t)^6\|
 D^6 E\|^2\leq c\Lambda_0,\ \ 1\leq k\leq5, \label{1.9}
 \egr
where the coefficient $c>0$ is independent of $\varepsilon$, and
 $\mathcal{H}^{k}(\mathbb{R}^3)$ denotes the space that
 $\{f\in L^6(\mathbb{R}^3),Df\in
H^{k-1}(\mathbb{R}^3)\},k\geq 1$. $D^kf$ denotes the $k $-times
spatial derivative of $f$}.
\end{theorem}

\begin{remark}  By $(\ref{1.8})$-$(\ref{1.9})$ and Nirenberg's inequality
for three dimensional case
 \be \| u\|_{L^\infty(\mathbb{R}^3)}\leq c\| D^2
u\|_{L^2(\mathbb{R}^3)}^{\frac{1}{2}}\|
u\|_{L^6(\mathbb{R}^3)}^{\frac{1}{2}}\leq c\| D^2
u\|_{L^2(\mathbb{R}^3)}^{\frac{1}{2}}\|
 Du\|_{L^2(\mathbb{R}^3)}^{\frac{1}{2}}\label{1.10}
  \ee
we can get the optimal $L^\infty$ time-decay rate of the solution
 \be
\|(\rho_i-\rho_i^*,u_i,E) \|_{L^\infty(\mathbb{R}^3)}\leq
c(1+t)^{-\frac{3}{4}}.\label{1.11}
 \ee
 This time-decay rate is the
same order as the heat equation in three dimension. In fact, when
taking relaxation limit for bipolar QHD, we can get the bipolar
quantum Drift-Diffusion (QDD) equation \qef{1.11'}-\qef{1.11''}
below. For this bipolar QDD model, we can show that the global
\textbf{solution of initial value problem tends to the equilibrium
state with the same rate as heat equation}~\cite{BQDD}.
 \be
  \partial_t\rho_i
 +\nabla\cdot[q_i\rho_iE-\nabla P_i(\rho_i)+\frac{\varepsilon
^2}{2}\rho_i\nabla(\frac{\triangle\sqrt{\rho_i}}{\sqrt{\rho_i}})]=0,\label{1.11'}
\ee \be \lambda^2\nabla\cdot E=\rho_a-\rho_b-\mathcal{C}(x),~\nabla
\times E =0,~E(x)\rightarrow 0,~|x|\rightarrow
+\infty.\label{1.11''}
  \ee
 \end{remark}
\textbf{Unlike the unipolar quantum hydrodynamical model}
\cite{JL2003,LM215-247,HLM2003,HLMO2003}, in Theorem~\ref{theorem
1.1} we can not get the exponential convergence to the asymptotical
equilibrium state for bipolar quantum model for the whole space case
due to the coupling and cancelation interaction between two
carriers. In fact, by the original equations \ef{1.1}--\ef{1.3}, we
can get the linearized system around the equilibrium state for the
variables
$$
(W_a,J_a,W_b,J_b,E)=
(\rho_a-\rho_a^*,\rho_au_a,\rho_b-\rho_b^*,\rho_bu_b,E)
$$
 that
 \be
 \left\{ \begin{array}{ll} W_{at}+\nabla\cdot J_a=0\\
J_{at}+P_a^{'}(\rho_a^*)\nabla
W_a-\frac{\varepsilon^2}{4}\nabla\triangle
W_a+J_a-\rho_a^*E=0\\
 W_{bt}+\nabla\cdot J_b=0\\
 J_{bt}+P_b^{'}(\rho_b^*)\nabla
W_b-\frac{\varepsilon^2}{4}\nabla\triangle
W_b+J_b+\rho_b^*E=0\\
\nabla\cdot E=W_a-W_b~~~~\nabla\times E=0,~~E\rightarrow 0
~~\text{as}~|x|\rightarrow \infty\end{array}\right. \label{1.12'}
\ee with initial data given by
 \be
(W_a,J_a,W_b,J_b)(x,0)=(W_{a0},J_{a0},W_{b0},J_{b0})(x)
\label{1.13'}
  \ee
   where we have let $\tau=1,\lambda=1$ for simplicity. From the Poisson equation $(\ref{1.12'})_5$ for
the electric potential $E$ we can represent $E$ by
 \be
E=\nabla\triangle^{-1}(W_a-W_b). \label{1.14'} \ee

Assume that the initial data (\ref{1.13'}) satisfies
 \be
J_{a0},J_{b0}\in H^5(\mathbb{R}^3),\quad W_{a0},W_{b0}\in
H^6(\mathbb{R}^3)\cap L^1(\mathbb{R}^3), \label{1.15'}
 \ee
so that the initial electric field $E_0$ obtained from Poisson
equation~\ef{1.14'} at initial time has the regularity
 \be
E_0=\nabla\triangle^{-1}(W_{a0}-W_{b0})\in H^5(\mathbb{R}^3).
 \ee
\begin{remark}
\textbf{The norm $\|D^kE\|$ of $E$ with integer $k>0$ can be
obtained by Lemma \ref{lemma 2.1} through the Poisson equation and
the norm $\|E\|_{L^2}$ is from the Riesz's potential theory in $R^3$
that $\|E\|\leq c \|(W_a-W_b)\|_{L^{\frac65}} $ with a positive
constant $c$.}
\end{remark}
For simplicity, we just consider the IVP~\ef{1.12'}--\ef{1.13'}
for following case
 \be
 \frac{\varepsilon^2}{4}=1,\ \rho_a^*=2, \ \rho_b^*=1,\,
c^*=1,\ P_a^{'}(2)=P_b^{'}(1)=1, \label{case}
 \ee
since the method used in section 4 to prove theorem~\ref{theorem
1.3} about the time-decay rate of solutions to
IVP~\ef{1.12'}--\ef{1.13'} can be applied to general case instead of
\ef{case}.

We have the algebraic time-decay rate of global solution to IVP
problem \ef{1.12'}--\ef{1.13'} for the case \ef{case} below.
\begin{theorem}{\label{theorem 1.3}}
Suppose  that \qef{1.15'}-\qef{case} hold. Assume that the Fourier
transformation $(\hat{W}_{a0},\hat{W}_{b0})$ of initial density
satisfy for some constants $m_0>0,\ r>0$ that
 \be \inf_{\xi\in
B(0,r)}|(\hat{W}_{a0}+2\hat{W}_{b0})(\xi)|\geq m_0, \label{1.16'}
 \ee
and the initial perturbation of momentum satisfies
 \be \nabla\cdot (J_{a0}+2J_{b0})=0.\label{1.17'}
 \ee
Then, the unique global solution to \qef{1.12'}--\qef{1.13'} exists
and satisfies
 \bgr
 W_{a},W_{b}\in C([0,+\infty),H^6(\mathbb{R}^3)),\
  J_{a},J_{b} \in C([0,+\infty),H^5(\mathbb{R}^3)),\nnm\\[2mm]
  E\in C([0,+\infty),H^5(\mathbb{R}^3)),\nnm
\egr
and
 \bma
&c_1(1+t)^{-\frac{k}{2}-\frac34}
 \leq
 \|(\partial_x^k W_a,\partial_x^kW_b)(t)\|_{L^2(\mathbb{R}^3)}
 \leq
c_2(1+t)^{-\frac{k}{2}},\ \ 0\leq k\leq6,  \label{1.18'}
\\
&c_1(1+t)^{-\frac{k}{2}-\frac54}
 \leq
 \|(\partial_x^kJ_a,\partial_x^kJ_b)(t)\|_{L^2(\mathbb{R}^3)}
 \leq
c_2(1+t)^{-\frac{k}{2}},\ \ 0\leq k\leq5 \label{1.19'}
 \ema
  for
$i=a,b$. The positive constants $c_1,c_2$ depend on $m_0$,
$\|U_0\|_{H^6\times H^5}$,  and $\|(W_{a0},W_{b0})\|_{L^1}$.
\end{theorem}

\begin{remark}
The theorem~\ref{theorem 1.3} shows that for above linearized
bipolar QHD, the density and momentum have only algebraic time-decay
rate from both above and below. This fact means that in general one
can only expect an algebraic time-decay rate for the original IVP
problem for nonlinear bipolar QHD \qef{1.1}--\qef{1.3}, since the
nonlinear bipolar QHD system can be viewed as a small perturbation
of the corresponding linearized system.
\end{remark}

As one can see that all the estimates \ef{1.8}--\ef{1.9} and
\ef{1.11} hold uniformly with respect to the Planck constant
$\varepsilon$, thus we can apply the theorem established in
\cite{ZLZ2006} to pass into the semiclassical limit
$\varepsilon\to0_+$ in \ef{1.1}--\ef{1.4}, and obtain the algebraic
time decay rate of the following limiting solution (which is the
solution of the limiting equation-- the classical bipolar
hydrodynamical model) as $\varepsilon\rightarrow 0_+$ below
 \bgr
\partial_t(\rho_i)+\nabla\cdot(\rho_i
u_i)=0,\label{1.12}\\
\partial_t(\rho_i u_i)+\nabla\cdot(\rho_iu_i \otimes u_i)+
\nabla P_i(\rho_i)=q_i\rho_iE-\frac{\rho_i
u_i}{\tau_i},\label{1.13}\\
\lambda^2\nabla\cdot E=\rho_a-\rho_b-\mathcal{C}(x),~\nabla \times E
=0,~E(x)\rightarrow 0,~|x|\rightarrow +\infty.
 \label{1.14} \egr

We have the following result about the decay rate of the
corresponding solution of bipolar HD model as an application of
Theorem $\ref{theorem 1.1}$ in the process of semiclassical limit.

 \begin{theorem}\label{corollary 1.3}
Under the assumptions of Theorem~\ref{theorem 1.1}, there exists
$(\rho_i,u_i,E)$, $i=a,b,$ such that as
$\varepsilon\rightarrow0_+$, the solution
$(\rho_i^{\varepsilon},u_i^{\varepsilon},E^{\varepsilon})$ of IVP
\ef{1.1}--\ef{1.4} tends to $(\rho_i,u_i,E)$ strongly
$$
 \rho_i^{\varepsilon} \rightarrow \rho_i ~~ in~~
C(0,T;C_b^3\cap H_{loc}^{5-s}); ~~u_i^{\varepsilon}\rightarrow
u_i~~ in~~ C(0,T;C_b^3\cap \mathcal{H}_{loc}^{5-s});$$
$$E^{\varepsilon}\rightarrow E~~ in~~
C(0,T;C_b^4\cap \mathcal{H}_{loc}^{6-s}),\quad s\in
(0,\mbox{$\frac{1}{2}$}).
$$
And where $(\rho_i,u_i,E)$ is the solution of the bipolar HD model
\ef{1.12}--\ef{1.14} with initial data \ef{1.4}. Moreover, it also
holds
  \be
\|(\rho_i-\rho_i^*,u_i,E) \|_{L^\infty(\mathbb{R}^3)}\leq
c(1+t)^{-\frac{3}{4}}\label{1.15}
 \ee
 as $t\rightarrow+\infty$.
 \end{theorem}

\par
\bigskip

The rest part of the paper is arranged as follows. After some
preliminary given in section 2, we shall prove Theorem 1.1 and
Theorem \ref{corollary 1.3} in the section 3, and we will prove
Theorem 1.3 in Section 4.
\section{ Some preliminary }
\setcounter{equation}{0} {\bf Notations } \par$C$ and $c$ always
denote the generic positive constants. $L^2(\mathbb{R}^3)$ is the
space of square integral functions on $\mathbb{R}^3$ with the norm
$\|\cdot\|$ or $\|\cdot\|_{L^2(\mathbb{R}^3)}$. $H^k(\mathbb{R}^3)$
with integer $k\geq 1$ denotes the usual Sobolev space of function
$f$ satisfying $\partial_x^i f\in L^2(\mathbb{R}^3)(0\leq i\leq k)$
with norm
$$\| f\|_k=\sqrt{\sum\limits _{\small{0\leq
|\alpha|\leq k}}\| D^\alpha f\|^2},$$ here and after $\alpha\in
N^3,D^\alpha =\partial_{x_1}^{s_1}\partial_{x_2}^{s
_2}\partial_{x_3}^{s _3}$ for $|\alpha |=s_1+s _2+s_3$, Especially
$\|\cdot\|_0=\|\cdot\|$. Let $\cal{B}$ be a Banach space,
$C^k([0,t];\cal{B})$ denotes the space of $\cal{B}$-valued k-times
continuously differentiable functions on [0,t]. We can extend the
above norm to the vector-valued function $u=(u_1,u_2,,u_3)$ with
$|D^\alpha u|^2=\sum \limits_{r=1}^3{|D^\alpha u_r|^2}$ and$$\|
D^ku\|^2=\int_{\mathbb{R}^3}(\sum\limits_{r=1}^3\sum\limits_{|\alpha|=k}(D^\alpha
u_r)^2)dx,$$ and $\| u\|_k=\| u\|_{H^k(\mathbb{R}^3)}=\sum
\limits_{i=0}^k\| D^iu\|$, $\| f\|_{L^\infty
([0,T];\cal{B})}=\sup\limits_{0\leq t\leq T}\| f(t)\|_{\cal{B}}$. We
also use the space $\mathcal{H}^{k}(\mathbb{R}^3)=\{f\in
L^6(\mathbb{R}^3),Df\in H^{k-1}(\mathbb{R}^3)\},k\geq 1$. Sometimes
we use $\| (.,.,...)\|_{H^k(\mathbb{R}^3)}$ or
 $\|(.,.,...)\|_k$ to denote the norm of the space
$H^k(\mathbb{R}^3)\times H^k(\mathbb{R}^3)\times\cdot\cdot\cdot
\times H^k(\mathbb{R}^3)$ and the $\mathcal{H}^{k}(\mathbb{R}^3)$ as
well.
 \begin{lemma}{\label{lemma 2.1}}
Let $f\in H^s(\mathbb{R}^3),s\geq\frac{3}{2}$. There is a unique
solution of the divergence equation \be\nabla\cdot
u=f,~~~\nabla\times u=0,~~~u(x)\rightarrow 0,~~~|x|
\rightarrow+\infty.\nnm\ee satisfying \be\| u
\|_{L^6(\mathbb{R}^3)}\leq C\| f \|_{L^2(\mathbb{R}^3)}, ~~~\|
Du\|_{H^s(\mathbb{R}^3)}\leq C\| f\|_{H^s(\mathbb{R}^3)}.\nnm\ee
 \end{lemma}

\par We will also use the Moser type calculus lemmas.
 \begin{lemma}
Let $f,g \in H^s(\mathbb{R}^3)\bigcap L^\infty(\mathbb{R}^3)$, then
it holds
 \bgr
 \| D^\alpha (fg)\|\leq C\| g\|_{L^\infty}\cdot\| D^\alpha
 f\|+C\| f\|_{L^\infty}\cdot\| D^\alpha
 g\|\nnm\\
\| D^\alpha (fg)-fD^\alpha g\|\leq C\| g\|_{L^\infty}\cdot\|
D^\alpha
 f\|+C\| f\|_{L^\infty}\cdot\|
 D^{|\alpha|-1}
 g\|\nnm
 \egr
 for $\alpha\in N^3,1\leq |\alpha|\leq s ,s\geq 0 $ is an
 integer.
  \end{lemma}
 \begin{lemma} Let $f\in H^s(\mathbb{R}^3)$ with $s\geq 0$ be an integer  and function $F(\rho)$ smooth enough and $F(0)=0$
then $F(f)(x)\in H^s(\mathbb{R}^3)$ and
$$\| F(f)\|_{H^s(\mathbb{R}^3)}\leq C\|
f\|_{H^s(\mathbb{R}^3)}.~$$

 \end{lemma}
\section{The proof of Theorem \ref{theorem 1.1} and Theorem \ref{corollary 1.3}}
\setcounter{equation}{0} Note that the local and global existence of
the solution in Theorem 1.1 can be referred to \cite{ZLZ2006}, we
only focus on the convergence rate of the solution to the
corresponding steady state.
\subsection{The reformulation of original problem}
Our idea is to obtain the uniform estimates of the local solution,
and we need to reformulate the original problem into a convenient
form. Take $\lambda=1$, $\tau=1$ and use $(.)_t$ to denote
$\partial_t(.)$ for convenience. First, by equations
\ef{1.1}--\ef{1.2} we can get the equations for
$\psi_i=\sqrt{\rho_i}\ ~(i=a,b)$ as in \cite{ZLZ2006}
 \bma
 \psi_{itt}+\psi_{it}+&\frac{\varepsilon^2\triangle^2\psi_i}{4}
 +\frac{q_i}{2\psi_i}\nabla\cdot(\psi_i^2E)
-\frac{1}{2\psi_i}\nabla^2(\psi_i^2 u_i\otimes u_i)
\nnm \\
-&\frac{1}{2\psi_i}\triangle
P_i(\psi_i^2)+\frac{\psi_{it}^2}{\psi_i}-\frac{\varepsilon^2|\triangle\psi_i|^2}{4\psi_i}=0,
\label{(3.1)}
 \ema
with the initial value
$$\psi_i(x,0):=\psi_{i_0}(x)=\sqrt{\rho_{i_0}(x)},$$
$$
\psi_{it}(x,0):=\psi_{i1}(x)=-\frac{1}{2}\psi_{i_0}\nabla\cdot
u_{i_0}-u_{i_0}\cdot
  \nabla\psi_{i_0}.$$
\textbf{By equation $(\ref{1.2})$ with the fact
$(u_i\cdot\nabla)u_i=\frac{1}{2}\nabla(|u_i|^2)-u_i\times(\nabla
\times u_i),$ taking curl of the two sides of the equation
$(\ref{1.2})$ we get for $\phi_i=\nabla\times u_i $ as
 \be
\phi_{it}+\phi_i+(u_i\cdot\nabla)\phi_i +\phi_i\nabla\cdot
u_i-(\phi_i\cdot\nabla)u_i-u_i(\nabla\cdot\phi)=0.\label{3.2}
 \ee
Here we have $\nabla\cdot\phi=0$. Introducing new variables
 $w_i=\psi_i-\sqrt{\rho_i^*}$
 then the system for} $(w_a,w_b,\phi_a,\phi_b,E)$ is
 \bgr
w_{att}+w_{at}+\frac{\varepsilon^2\triangle^2w_a}{4}+\frac{1}{2}(w_a+\sqrt{\rho_a^*})
\nabla\cdot E-P_a^\prime (\rho_a^*)\triangle
w_a=f_{a1},\label{3.3}\\
  w_{btt}+w_{bt}+\frac{\varepsilon^2\triangle^2w_b}{4}-\frac{1}{2}(w_b+\sqrt{\rho_b^*})
\nabla\cdot E-P_b^\prime (\rho_b^*)\triangle
w_b=f_{b1},\label{3.4}\\
 \phi_{at}+\phi_a=f_{a2},\label{3.5}\\
 \phi_{bt}+\phi_b=f_{b2},\label{3.6}\\
 \nabla\cdot
E=w_a^2-w_b^2+2\sqrt{\rho_a^*}w_a-2\sqrt{\rho_b^*}w_b,~~\nabla\times
E=0,
 \label{3.7}
  \egr
with the initial conditions given by
 \bma
&w_i(x,0):=w_{i_0}(x)=\psi_{i_0}-\sqrt{\rho_i^*},\
\phi_i(x,0):=\phi_{i_0}(x)=\nabla \times
u_{i_0}(x),\label{3.7'}\\
&w_{it}(x,0):=w_{i1}(x)=(-u_{i_0}\cdot\nabla w_{i_0}
-\frac{1}{2}(w_{i_0}+\sqrt{\rho_i^*})\nabla\cdot
u_{i_0})\label{3.8'}
 \ema
\textbf{and where
  \bma
 f_{i1}:=f_{i1}(x,t) = &\frac{-w_{it}^2}{w_i+\sqrt{\rho_i^*}}-q_i\nabla w_iE
  +(P_i^\prime((w_i+\sqrt{\rho_i^*})^2)-P_i^\prime(\rho_i^*))\triangle w_i\nnm\\
  &+2(w_i+\sqrt{\rho_i^*})P_i^{\prime\prime}((w_i+\sqrt{\rho_i^*})^2)|\nabla
w_i|^2+P_i^\prime((w_i+\sqrt{\rho_i^*})^2)\frac{|\nabla
w_i|^2}{w_i+\sqrt{\rho_i^*}}\nnm\\
&+\frac{\varepsilon^2(\triangle w_i)^2}{4(w_i+\sqrt{\rho_i^*})}
   +\frac{\nabla^2((w_i+\sqrt{\rho_i^*})^2u_i\otimes
   u_i)}{2(w_i+\sqrt{\rho_i^*})},\label{3.8}\\
f_{i2}:=f_{i2}(x,t)=&((\phi_i\cdot\nabla)u_i-(u_i\cdot\nabla)\phi_i-\phi_i\nabla\cdot
u_i),\label{3.9}
  \ema
for $i=a,b$. 
 %
We will also use the relation between}
 $\nabla\cdot
u_i$ and $\nabla w_i,w_{it}$ from (1.2)
 \be
 2w_{it}+2u_i\cdot\nabla
w_{i}+(w_{i}+\sqrt{\rho_i^*})\nabla\cdot u_i=0.\label{3.10}
 \ee
\subsection{The a-priori estimates  }
\par
Assume that the classical solutions $w_i,u_i,E$ satisfy a-priorily
 \bma
 \delta_T\stackrel{\triangle}{=}\max_{0\leq t\leq T}
     &\{\sum_{k=0}^{5}(1+t)^k\| D^k w_i\|^2
 +\sum_{k=1}^{5}(1+t)^k\| D^k u_i\|^2
 +\sum_{k=0}^{3}(1+t)^{k+2}\| D^k w_{it}\|^2
 \nnm\\
 &+\sum_{k=1}^{3}(1+t)^{2+k}\| D^k
 u_{it}\|^2+(1+t)^5\| D^4 w_{it}\|^2
 \nnm\\
 &+\sum_{k=1}^{5}(1+t)^{k}\| D^{k} E\|^2
 +\sum_{k=0}^{2}(1+t)^{3+k}\| D^k
 w_{itt}\|^2\}\label{3.12}\ll1.
 \ema
It follows for the sufficiently small $\delta_T$ the positivity of
density $\psi_i\ (i=a,b)$ as
$$
\frac{\sqrt{\rho_i^*}}{2}\leq
w_i+\sqrt{\rho_i^*}\leq\frac{3}{2}\sqrt{\rho_i^*}.
$$
By Nirenberg's inequality for three-dimensional case from \ef{3.12},
we
 have
 \bma &\sum_{k=0}^{3}(1+t)^{k+1}\| D^k w_i\|_{L^\infty}^2+\sum_{k=0}^{2}(1+t)^{k+3}\|
 D^k
w_{it}\|_{L^\infty}^2+(1+t)^4\| w_{itt}\|_{L^\infty}^2\leq
c\delta_T\label{3.13}\\
 &\sum_{k=0}^{3}(1+t)^{k+1}\|
D^k u_i\|_{L^\infty}^2+\sum_{k=0}^{1}(1+t)^{k+3}\| D^k
u_{it}\|_{L^\infty}^2+\sum_{k=0}^{3}(1+t)^{k+1}\| D^k
E\|_{L^\infty}^2\leq c\delta_T.\label{3.14}
 \ema
\par

With the help of the a-priori assumptions \ef{3.12} we establish the
following a-priori estimates
\begin{lemma}
For the short time solution $(w_i,u_i,E)$ it holds for $t\in[0,T]$
 that
 \bma
 &\sum_{k=0}^5(1+t)^k\| D^k w_i\|^2+(1+t)^5\|
\varepsilon D^6 w_i\|^2+\sum_{k=0}^3(1+t)^{k+2}\| D^k w_{it}\|^2
\nnm\\
  &+(1+t)^5\| D^4 w_{it}\|^2+\sum_{k=0}^2(1+t)^{3+k}\| D^k
w_{itt}\|^2\leq
  c\Lambda_0,\label{3.16}\\
&\sum_{k=1}^5(1+t)^k\| D^k u_i\|^2+\sum_{k=1}^3(1+t)^{2+k}\| D^k
u_{it}\|^2\leq c\Lambda_0,\label{3.17}\\
& \sum_{k=1}^5(1+t)^k\| D^k E\|^2+\int_0^t \sum_{k=1}^5(1+s)^{k-1}\|
D^k E\|^2ds\leq c\Lambda_0,\label{3.18}
 \\
 & \int_0^t\{\sum_{k=1}^5(1+s)^{k-1}\| D^k w_i\|^2
+\sum_{k=0}^4(1+s)^{k+1}\| D^k w_{it}\|^2\}ds\leq c\Lambda_0,\label{3.19}\\
&\int_0^t\{\sum_{k=1}^5(1+s)^k\| D^k u_i\|^2
+\sum_{k=1}^3(1+s)^{k+2}\| D^k u_{it}\|^2\}ds\leq
c\Lambda_0,\label{3.20}
 \ema
provided $\delta_T$ is small enough, where the $\Lambda_0$ is
 defined in Theorem~1.1.
\end{lemma}
\textbf{Proof}: \ {\it Step 1 (the basic estimates)}. Multiplying
equation $(\ref{3.3})$ by $(w_a+2w_{at})$, and (\ref{3.4}) by
$(w_b+2w_{bt}),$ integrating by parts the resulted equations over
$\mathbb{R}^3$, omitting $\mathbb{R}^3$ without confusion, summing
the resulted two equalities and noticing the fact from Poisson
equation (\ref{3.7}) that
 \bma
&\int\{(\frac{1}{2}(w_a+\sqrt{\rho_a^*})\nabla\cdot
E)(w_a+2w_{at})
   -(\frac{1}{2}(w_b+\sqrt{\rho_b^*})\nabla\cdot E)(w_b+2w_{bt})
   \}dx\nnm\\
   &=\frac{1}{4}\frac{d}{dt}\int|\nabla\cdot E|^2dx+\frac{1}{4}\int_{R^3}|\nabla\cdot
   E|^2dx-\frac{1}{4}\int_{R^3}\nabla(w_a^2-w_b^2)\cdot E
   dx\nnm
 \ema
we can get
 \bma
 \frac{d}{dt} \int\{w_{at}^2+ &w_aw_{at}+\frac{w_a^2}{2}
+w_{bt}^2+ w_bw_{bt}+\frac{w_b^2}{2}+ P_a^\prime(\rho_a^*)|\nabla
w_a|^2+P_b^\prime(\rho_b^*)|\nabla
w_b|^2\nnm\\
   &+\frac{\varepsilon^2}{4}(|\triangle w_a|^2
   +|\triangle w_b|^2)+\frac{1}{4}|\nabla\cdot E|^2\}dx
   \nnm\\
   + \int\{(w_{at}^2&+w_{bt}^2)
   + P_a^\prime(\rho_a^*)|\nabla w_a|^2+ P_b^\prime(\rho_b^*)|\nabla w_b|^2+\frac{\varepsilon^2}{4}(|\triangle w_a|^2
   +|\triangle w_b|^2)\nnm\\
   &+\frac{1}{4}|\nabla\cdot E|^2\}dx\nnm\\
  =\frac{1}{4}\int(\nabla&(w_a^2-
   w_b^2)\cdot E dx\nnm\\
   +\int\{&f_{a1}(x,t)(w_a+2w_{at})
   +f_{b1}(x,t)(w_a+2w_{bt})\}dx .\label{3.21}
   \ema

 \textbf{By assumptions \ef{3.12}, using
Sobolev imbedding theorem and H\"{o}lder's inequality, Young's
inequality and integration by parts, we can estimate the
right-hand side terms of (\ref{3.21}) as follows
  \bma \int w_i\nabla w_i\cdot
E dx\leq& \| w_i\|_{L^3}\|\nabla w_i\|_{L^2}\| E \|_{L^{6}}
\nnm\\
\leq &c(\| w_i\|_{L^2}+\| \nabla w_i\|_{L^2}) \| \nabla
w_i\|_{L^2}\cdot\|  E
\|_{L^6}\nnm\\
\leq &c\delta_T(\| \nabla w_i\|^2+\| \nabla\cdot E
\|^2),\label{3.22}
 \ema
 and
  \bma
&\int[P_i^\prime((w_i+\sqrt{\rho_i^*})^2)-P_i^\prime(\rho_i^*)]
\triangle w_i\cdot(2w_{it})dx\nnm\\
 \leq& -\frac{d}{dt} \int[P_i^\prime((w_i+\sqrt{\rho_i^*})^2)-P_i^\prime(\rho_i^*)]|\nabla w_i|^2dx
 +c\delta_T\|(\nabla w_i,w_{it})\|^2,\label{3.23}\\
 &\int u_i\cdot\nabla w_{it}(2w_{it})dx
=-\int\nabla\cdot u_i(w_{it})^2dx\leq c\delta_T\|
w_{it}\|^2,\label{3.24}\\
&\int u_i\nabla(u_i\cdot\nabla w_i )\cdot2w_{it}dx
\leq-\frac{d}{dt}\int( u_i\cdot\nabla w_i)^2dx +c\delta_T\|(\nabla
w_i,w_{it})\|^2 ,\label{3.26}
 \ema
where we have used the fact} $\| Du_i\|^2\leq c(\|\nabla\cdot
u_i\|^2+\|\nabla\times u_i\|^2),$ and $\|\nabla w_i\|^2,\|
w_{it}\|^2$ to estimate $\nabla\cdot u_i$ through equation
\ef{3.10}. The other terms in the right-hand side of (\ref{3.21})
can also be estimated easily by integration by parts, H\"{o}lder's
inequality, Young's inequality and the Lemma 2.2 and Lemma 2.3,
together with $(\ref{3.22})$-$(\ref{3.26})$ we can have from
\ef{3.21} that
 \bma
 \frac{d}{dt} \int\{&w_{at}^2+ w_aw_{at}+\frac{w_a^2}{2}
+w_{bt}^2+ w_bw_{bt}+\frac{w_b^2}{2}+ P_a^\prime(\rho_a^*)|\nabla
w_a|^2+P_b^\prime(\rho_b^*)|\nabla
w_b|^2\nnm\\
   &+\frac{\varepsilon^2}{4}(|\triangle w_a|^2
   +|\triangle w_b|^2)+\frac{1}{4}|\nabla\cdot E|^2
   +[P_a^\prime((w_a+\sqrt{\rho_a^*})^2)-P_a^\prime(\rho_a^*)]|\nabla w_a|^2  \nnm\\
  &+[P_b^\prime((w_b+\sqrt{\rho_a^*})^2)-P_b^\prime(\rho_b^*)]|\nabla w_b|^2
  +( u_a\cdot\nabla w_a)^2+( u_b\cdot\nabla w_b)^2\}dx\nnm\\
 + \int\{(w_{at}^2&+w_{bt}^2)
   + P_a^\prime(\rho_a^*)|\nabla w_a|^2+ P_b^\prime(\rho_b^*)|\nabla w_b|^2+\frac{\varepsilon^2}{4}(|\triangle w_a|^2
   +|\triangle w_b|^2)+\frac{1}{4}|\nabla\cdot E|^2\}dx\nnm\\
\leq c\delta_T\|(\nabla &w_a,\nabla w_b,w_{at},w_{bt},\nabla\cdot
E,\phi_a,\phi_b)\|^2.\label{3.27}
   \ema

Taking inner product between $(\ref{3.5})$ and $2\phi_a$, and
between $(\ref{3.6})$ and $2\phi_b$, integrating over
$\mathbb{R}^3$, we obtain
 \be \frac{d}{dt}\int (|\phi_a|^2+|\phi_b|^2)dx
+2\int (|\phi_a|^2+|\phi_b|^2)dx=\int
\{f_{a2}\cdot2\phi_a+f_{b2}\cdot2\phi_b\}dx.\label{3.28}
 \ee A
simple analysis to the right-hand side of \ef{3.28} together with
\ef{3.12} gives
 \be \frac{d}{dt}\int
(|\phi_a|^2+|\phi_b|^2)dx +2\int (|\phi_a|^2+|\phi_b|^2)dx\leq
c\delta_T\|(\phi_a,\phi_b,\nabla w_a,\nabla
w_b,w_{at},w_{bt})\|^2.\label{3.29}
 \ee

Integrating of the summation of $(\ref{3.27})$ and $(\ref{3.29})$
over $[0,t]$ and using
$$P_i^\prime(\rho_i^*)>0,\ \ \frac{1}{6}(x^2+y^2)\leq
x^2+xy+\frac{y^2}{2}\leq2(x^2+y^2),$$ we obtain
 \bma
  &\|(w_a,w_b)\|_1^2+\|(\varepsilon D^2w_a,\varepsilon D^2w_b)\|^2
  +\|(w_{at},w_{bt},
   \phi_a,\phi_b,D
   E)\|^2\nnm\\
 &+\int_0^t \{\|(\nabla w_a,\nabla w_b
 ,\varepsilon D^2 w_a,\varepsilon D^2 w_b,w_{at},w_{bt},\phi_a,\phi_b)\|^2+\|D
   E\|^2\}ds\nnm\\
   &\leq c\Lambda_0.
    \label{3.33''}\ema
\textbf{Making summation between the integral
$\int\{(\ref{3.3})\times2(1+t)w_{at}+(\ref{3.4})\times2(1+t)w_{bt}\}dx$
and  $\ef{3.28}\times(1+t)$, we can have after a complicated but
straightforward computation that
 \bma
&\frac{d}{dt}\{(1+t)\int\{w_{at}^2+w_{bt}^2+
P_a^\prime(\rho_a^*)|\nabla w_a|^2+P_b^\prime(\rho_b^*)|\nabla
w_b|^2+\frac{\varepsilon^2}{4}(|\triangle w_a|^2
   +|\triangle w_b|^2)\nnm\\
   &\ \ \ \ \ \ \ \ \ \ \ \ \ \ \ \
   \ \ \ \ \ +\frac{1}{4}|\nabla\cdot E|^2+|\phi_a|^2+|\phi_b|^2
   +[P_a^\prime((w_a+\sqrt{\rho_a^*})^2)-P_a^\prime(\rho_a^*)]|\nabla w_a|^2\nnm\\
   &\ \ \ \ \ \ \  \
   \ \ \ \ \ \ \ \ \ \ \ \ \ \
   +[P_b^\prime((w_b+\sqrt{\rho_b^*})^2)-P_b^\prime(\rho_b^*)]|\nabla
w_b|^2+ (u_a\cdot w_a)^2+(u_b\cdot w_b)^2 \}dx\}\nnm\\
  &\ \ \ +2(1+t)\|(w_{at},w_{bt},\phi_a,\phi_b)\|^2\nnm\\
  &\leq(1+t)\|(w_{at},w_{bt},\phi_a,\phi_b)\|^2+c\delta_T
  \|(\nabla w_a,\nabla w_b,\varepsilon D^2 w_a,\varepsilon D^2w_b,\phi_a,\phi_b,D E)\|^2\label{3.34'}
   \ema
where we have used the a-priori time-decay rate assumptions
\ef{3.12}, H\"{o}lder's inequality, Young's inequality to estimate
the right-hand side terms as follows}
   \bma
   &\frac{1}{4}\int (1+t)\nabla( w_a^2- w_b^2 )\cdot E dx\nnm\\
   &+\int(1+t)\{(f_{a1}(x,t)(w_a+2w_{at})
   +f_{b1}(x,t)(w_b+2w_{bt})\}dx\nnm\\
   &+\int
(1+t)\{f_{a2}\cdot2\phi_a+f_{b2}\cdot2\phi_b\}dx\nnm\\
&\leq(1+t)\|(w_{at},w_{bt},\phi_a,\phi_b)\|^2+c\delta_T
  \|(\nabla w_a,\nabla w_b,\varepsilon D^2w_a,\varepsilon D^2w_b,\phi_a,\phi_b,DE)\|^2.
   \nnm\ema
The integrating of \ef{3.34'} over $[0,t]$ together with the help
of \ef{3.33''} gives rise to
 \bma
  &(1+t)\|(\nabla w_a,\nabla w_b,\varepsilon D^2 w_a,\varepsilon D^2 w_b
  ,w_{at},w_{bt}
  ,\phi_a,\phi_b,D
   E)\|^2\nnm\\
 &+\int_0^t(1+s) \{\|(w_{at},w_{bt},
   \phi_a,\phi_b)\|^2ds\nnm\\
   &\leq c\Lambda_0.
    \label{3.35'}
    \ema
The combination of \ef{3.33''} and \ef{3.35'} shows the basic
estimates in Lemma 3.1 as
  \bma
&\| w_i\|^2+(1+t)\| D w_i\|^2
+(1+t)\| DE\|^2+(1+t)\| Du_i\|^2\leq c\Lambda_0,\label{3.36'}\\
&\int_0^t\|(\nabla w_i,DE)\|^2ds+\int_0^t(1+s)(\| Du_i\|^2+\|
w_{it}\|^2)ds\leq c\Lambda_0,\label{3.38'}
 \ema
{\it Step 2 (the higher order estimates).} Next, we will do the
higher order estimates. To this end, set $\widetilde{w_i}:=D^\alpha
w_i,\ \widetilde{\phi_i}:=D^\alpha \phi_i,\ \widetilde{E}:=D^\alpha
E(i=a,b.\ 1<|\alpha|\leq4)$. Differentiating equations
\ef{3.3}--\ef{3.7} with respect to $x$, we get the equations for
$\widetilde{w_i},\widetilde{\phi_i},\widetilde{E}$ that
 \bma
&\widetilde{w_i}_{tt}+\widetilde{w_i}_{t}+\frac{\varepsilon^2}{4}
\triangle^2\widetilde{w_i}-P_i^\prime(\rho_i^*)\triangle\widetilde{w_i}
+ \frac{q_i}{2}(w_i+\sqrt{\rho_i^*})\nabla\cdot\widetilde{E}
\nnm\\
=&D^\alpha f_{i1}(x,t)-D^\alpha(\frac{q_i}{2}(w_i+\sqrt{\rho_i^*}
) \nabla\cdot E) +\frac{q_i}{2}(w_i+\sqrt{\rho_i^*}
)\nabla\cdot\widetilde{E},\label{3.40}
 \ema
 \be
\widetilde{\phi_i}_t+\widetilde{\phi_i}=D^{\alpha}
f_{i2},\label{3.41} \ee
\be \nabla\cdot\widetilde{E}=
D^{\alpha}(w_a^2-w_b^2+2\sqrt{\rho_a^*}w_a-2\sqrt{\rho_b^*}w_b),
\label{3.42}
 \ee
 $~ q_a=1,q_b=-1.$

\par

\textbf{Similarly to deriving the previous basic estimates, combining the
following integrals together
 \be
 \int_0^t\int\sum_{l=0}^{|\alpha|}\{\ef{3.40}_{i=a}\times(1+s)^l(D^{\alpha}{w_a}+2D^{\alpha}{w_a}_{t})
+\ef{3.40}_{i=b}\times(1+s)^l(D^{\alpha}{w_b}+2D^{\alpha}{w_b}_{t})\}dxds\nnm
\ee
 \be \int_0^t\int
 \sum_{l=0}^{|\alpha|}\{\ef{3.41}_{i=a}\cdot2(1+s)^lD^{\alpha}{\phi_a}
+\ef{3.41}_{i=b}\cdot2(1+s)^lD^{\alpha}{\phi_b}\}dxds\nnm
 \ee
 and
  \be
\int_0^t\int\{\ef{3.40}_{i=a}\times2(1+s)^{|\alpha|+1}D^{\alpha}{w_a}_{t}
+\ef{3.40}_{i=b}\times2(1+s)^{|\alpha|+1}D^{\alpha}{w_b}_{t}\}dxds
\nnm
 \ee
  \be
\int_0^t\int\{\ef{3.41}_{i=a}\cdot2(1+s)^{|\alpha|+1}D^{\alpha}{\phi_a}
+\ef{3.41}_{i=b}\cdot2(1+s)^{|\alpha|+1}D^{\alpha}{\phi_b}\}dxds\nnm
  \ee
  for $|\alpha|=k$ with $k=1,2,3,4$ respectively, we can
get after a straightforward computation that}
 \bma
  &(1+t)^{k+1}\|(D^{k+1} w_a,D^{k+1} w_b,\varepsilon D^{k+2}w_a,\varepsilon D^{k+2}w_b
  ,D^kw_{at},D^kw_{bt},D^k\phi_a,D^k\phi_b,D^{k+1}
   E)\|^2\nnm\\
   &+\int_0^t(1+s)^{k} \|(D^{k+1}w_a,D^{k+1}w_b,\varepsilon D^{k+2}w_a,\varepsilon D^{k+2}w_b,
   D^{k+1}
   E)\|^2ds\nnm\\
 &+\int_0^t(1+s)^{k+1} \|(D^kw_{at},D^kw_{bt},D^k\phi_a,D^k\phi_b)\|^2ds\leq c\Lambda_0.
    \label{3.43}
    \ema
\textbf{By \ef{3.43} we can get part of decay rates in Lemma 3.1 that
 \bma
 &(1+t)^k\| D^k
w_i\|^2+(1+t)^5\|\varepsilon D^6w_i\|^2\leq
c\Lambda_0,&0\leq k\leq5,\label{3.44}\\
 &(1+t)^k\| D^k u_i\|^2+(1+t)^{k}\|
D^{k}E\|^2\leq c\Lambda_0,
&1\leq k\leq5,\label{3.45}\\
&\int_0^t\{(1+s)^{k-1}\|(D^k w_i,D^k E)\|^2+(1+s)^k\| D^k
u_i\|^2\}ds\leq c\Lambda_0,&1\leq k\leq5,\label{3.46}
 \ema
 and
 \bma (1+t)^{1+k}\| D^k w_{it}\|^2
+\int_0^t(1+s)^{1+k}\| D^k w_{it}\|^2ds\leq c\Lambda_0,\ \ 1\leq
k\leq4.\label{3.47}
 \ema
The higher order estimate $(1+t)^6\|D^6E\|^2\leq c\Lambda_0$ can be
obtained by Poisson equation \ef{3.7} and the Lemma 2.1.}

To complete the proof we still need to do the decay rate of
$(w_a,w_b,u_a,u_b)$ about higher order derivatives on time $t$. Set
$\bar{w_i}=D^{\alpha}w_{it},\ \bar{\phi_i}=D^{\alpha}\phi_{it},\
\bar{E}=D^{\alpha}E_t\ (0\leq|\alpha|\leq2)$, then we get the
equations for $\bar{w_i},\bar{\phi_i},\bar{E}$
 \bma
&\bar{w_i}_{tt}+\bar{w_i}_{t}+\frac{\varepsilon^2}{4}
\triangle^2\bar{w_i}+
\frac{q_i}{2}(w_i+\sqrt{\rho_i^*})\nabla\cdot\bar{E}
-P_i^\prime(\rho_i^*)\triangle\bar{w_i}\nnm\\
&=D^\alpha
(f_{i1}(x,t))_t-D^\alpha(\frac{q_i}{2}(w_i+\sqrt{\rho_i^*} )
\nabla\cdot E)_t+\frac{q_i}{2}(w_i+\sqrt{\rho_i^*}
)\nabla\cdot\bar{E},\label{3.48}\\
&\bar{\phi_i}_t+\bar{\phi_i}=D^{\alpha}(f_{i2}(x,t))_t,
\label{3.49}\\
&\nabla\cdot\bar{E}=D^{\alpha}
(w_a^2-w_b^2+2\sqrt{\rho_a^*}w_a-2\sqrt{\rho_b^*}w_b)_t,\label{3.50}
 \ema
 with $i=a,b,~ q_a=1,q_b=-1.$

Based on the results derived in \ef{3.44}--\ef{3.47} we can get from
\ef{3.48}--\ef{3.50} the more faster time-decay rate for
$\bar{w_i},\bar{\phi_i},\bar{E}$ as before. Summing the integrals
\bgr
\int_0^t\int\sum_{l=0}^{2+|\alpha|}\{(\ref{3.48})_{i=a}(1+s)^l(D^{\alpha}{w_a}_t+2D^{\alpha}{w_a}_{tt})
+(\ref{3.48})_{i=b}(1+s)^l(D^{\alpha}{w_b}_t+2D^{\alpha}{w_b}_{tt})\}dxds
\nnm
\\
\int_0^t\int\sum_{l=0}^{2+|\alpha|}
\{(\ref{3.49})_{i=a}\cdot2(1+s)^lD^{\alpha}{\phi_a}_t
+(\ref{3.49})_{i=b}\cdot2(1+s)^lD^{\alpha}{\phi_b}_t\}dxds \nnm
 \egr
and
 \bma
\int_0^t\int&\{(\ref{3.48})_{i=a}\times2(1+s)^{|\alpha|+3}D^{\alpha}{w_a}_{tt}+
(\ref{3.48})_{i=b}\times2(1+s)^{|\alpha|+3}D^{\alpha}{w_b}_{tt}\nnm\\
&\ \ \ \
+(\ref{3.49})_{i=a}\cdot2(1+s)^{|\alpha|+3}D^{\alpha}{\phi_a}_t
+(\ref{3.49})_{i=b}\cdot2(1+s)^{|\alpha|+3}D^{\alpha}{\phi_b}_t\}dxds
\nnm
 \ema
for $\alpha$ with $|\alpha|=0,1,2$ respectively which together
with the help of the results \ef{3.44}--\ef{3.47} gives us finally
 \bma
&(1+t)^{k+2}\| D^kw_{it}\|^2+\int_0^t(1+s)^{1+k}\| D^kw_{it}\|^2
ds\leq c\Lambda_0,& 0\leq k\leq3,\label{3.51}\\
 &(1+t)^{k+3}\|
D^kw_{itt}\|^2+\int_0^t(1+s)^{3+k}\| D^kw_{itt}\|^2 ds\leq
c\Lambda_0,& 0\leq
k\leq2,\label{3.52}\\
&(1+t)^{k+3}\| D^k\phi_{it}\|^2+\int_0^t(1+s)^{3+k}\|
D^k\phi_{it}\|^2 ds\leq c\Lambda_0,& 0\leq
k\leq2,\label{3.53}\\
&(1+t)^{k+2}\| D^kE_{t}\|^2+\int_0^t(1+s)^{1+k}\| D^kE_{t}\|^2
ds\leq c\Lambda_0,& 1\leq k\leq3.\label{3.54}
 \ema
 Note that
 $\| Du_t\|^2\leq c(\|\nabla\cdot u_t\|^2)+\| \nabla\times
u_t\|^2$ with the help of \ef{3.51}--\ef{3.54} and the relation of
$\nabla\cdot u_i$ and $\nabla w_i,\ w_{it}$ through the equation
\ef{3.10}, we have
 \bma (1+t)^{2+k}\|
D^ku_{it}\|^2+\int_0^t(1+s)^{2+k}\| D^ku_{it}\|^2ds\leq c\Lambda_0,\
\ 1\leq k\leq3.\label{3.55}
 \ema
Then, Lemma 3.1 follows from \ef{3.51}--\ef{3.55} and
\ef{3.44}--\ef{3.47} and \ef{3.36'}--\ef{3.38'}.\hfill $\square$

\subsection{The proof of main results }
\textsl{\textbf{Proof of Theorem $\ref{theorem 1.1}$ and
Theorem $\ref{corollary 1.3}$:}} \\
From Lemma 3.1, we know that the sufficiently small $\Lambda_0$
makes us be able to extend the solution to the global one by
continuity argument and the estimates \ef{3.16}--\ef{3.20} hold for
any $t>0$ especially that
 \bma
  &(1+t)^k\|
D^kw_i\|^2+(1+t)^5\|\varepsilon D^6w_i\|^2\leq c\Lambda_0,\ \ \
&0\leq
k\leq5,\label{3.56}\\
 & (1+t)^{k+2}\|
D^kw_{it}\|^2+(1+t)^5\| D^4w_{it}\|^2\leq c\Lambda_0,\ \ \ & 0\leq
k\leq3,\label{3.57}\\
& (1+t)^{k}\| D^ku_i\|^2+(1+t)^{k}\| D^kE\|^2\leq c\Lambda_0,\ \ \
&
1\leq k\leq5,\label{3.58} \\
& (1+t)^{2+k}\| D^ku_{it}\|^2\leq c\Lambda_0,&1\leq
k\leq3.\label{3.59}
 \ema
  \textbf{The coefficient $c$ is independent of the
Planck constant $\varepsilon$ and time $t$. As
$\rho_i=(w_i+\sqrt{\rho^*})^2$ we can get the conclusion of the
Theorem $\ref{theorem 1.1}$ that
 \bma &(1+t)^k\|
D^k(\rho_i-\rho_i^*)\|^2+(1+t)^5\| \varepsilon
D^6(\rho_i-\rho_i^*)\|^2\leq c\Lambda_0,\ \ \ &0\leq k\leq5,
\label{3.60}\\
 &(1+t)^k\| D^k u_i\|^2+(1+t)^k\|
 D^k E\|^2\leq c\Lambda_0,\ \ \ &1\leq
k\leq5.\label{3.61}
 \ema
 Thus, the proof of  Theorem $\ref{theorem
1.1}$ is completed. From \ef{3.60}--\ef{3.61}, using Nirenberg's
inequality we have}
 \be \|(\rho_i-\rho_i^*,u_i,E)
\|_{L^\infty(\mathbb{R}^3)}\leq c(1+t)^{-\frac{3}{4}}.\label{3.62}
 \ee
\par Let us turn to the proof of the Theorem~$\ref{corollary 1.3}$.
Since all above a-priori estimates established for the solutions
given in Theorem $\ref{theorem 1.1}$ hold uniformly with respect to
Planck constant $\varepsilon$. Denote the solution by
$(\rho_i^\varepsilon,u_i^\varepsilon,E^\varepsilon)$ and it follows
that(see\cite{ZLZ2006}) there is a solution denoted by
$(\hat{\rho_a},\hat{u_a},\hat{\rho_b},\hat{u_b},\hat{E})$ such that
$$
 \rho_i^\varepsilon \rightarrow \hat{\rho_i} ~~ in~~
C(0,T;C_b^3\cap H_{loc}^{5-s}); ~~u_i^\varepsilon\rightarrow
\hat{u_i}~~ in~~ C(0,T;C_b^3\cap \mathcal{H}_{loc}^{5-s});$$
$$E^\varepsilon\rightarrow \hat{E}~~ in~~
C(0,T;C_b^4\cap \mathcal{H}_{loc}^{6-s}),\quad s\in
(0,\mbox{$\frac{1}{2}$}),
$$
for any $T>0, i=a,b.$ One can easily verify that
$(\hat{\rho_a},\hat{u_a},\hat{\rho_b},\hat{u_b},\hat{E})$ is the
global-in time solution of the bipolar hydrodynamic model
$(\ref{1.12})$-$(\ref{1.14})$. What's more, we have the estimate by
\ef{3.62} that
 \be
\|(\hat{\rho_i}-\rho_i^*,\hat{u_i},\hat{E})
\|_{L^\infty(\mathbb{R}^3)}\leq c(1+t)^{-\frac{3}{4}}.\label{3.63}
 \ee
\par

\section{Algebraic decay rate for linearized system}
\label{sec4} \setcounter{equation}{0}
In this section, we will prove Theorem \ref{theorem 1.3}. Namely, we
shall show that for linearized bipolar QHD system, the density and
momentum converge to its asymptotical state at an algebraic decay
rate from both above and below. This implies that in general we can
only get an algebraic time-decay rate for bipolar QHD. This is
caused by the interactions between two carriers. Since the nonlinear
bipolar QHD system is a small perturbation of the corresponding
linearized system, one can only expect the similar results for the
original problem.\par

By (\ref{case}), the equation \ef{1.12'} for $U=(W_a,J_a,W_b,J_b)$
can be rewritten as
 \be
\left\{ \begin{array}{ll} W_{at}+\nabla\cdot J_a=0\\
J_{at}+\nabla W_a-\nabla\triangle
W_a+J_a-2\nabla\triangle^{-1}(W_a-W_b)=0\\
 W_{bt}+\nabla\cdot J_b=0\\
 J_{bt}+\nabla
W_b-\nabla\triangle W_b+J_b+\nabla\triangle^{-1}(W_a-W_b)=0
\end{array}\right.
\label{4.2}
\ee
with initial data given by
\be
U(x,0)=U_0(x)=:(W_{a0},J_{a0},W_{b0},J_{b0})(x). \label{4.2'}
 \ee

Let us write the solution of the linear problem
(\ref{4.2})--(\ref{4.2'}) formally
 \be
U=e^{At}U_0 \label{4.2''}\ee where $U$ will be the inverse
 of its Fourier transformation
$\hat{U}=(\hat{W}_a,\hat{J}_a,\hat{W}_b,\hat{J}_b)$
 whose equation can be derived by taking Fourier transform with respect to $x$
  on $(\ref{4.2})$ as
\be \left\{ \begin{array}{ll}\hat{U}_t=\hat{A} \hat{U}\\
\hat{U}(\xi,0)=(\hat{W}_{a0},\hat{J}_{a0},\hat{W}_{b0},\hat{J}_{b0})\end{array}\right.
\label{4.3}\ee
where the Matrix
\[
\hat{A}=\left( \begin{array}{cccc}
0&-\textbf{i}\xi^t&0&0\\[2mm]
-\textbf{i}\xi b_1&-I_{3}&\textbf{i}\xi d_1&0\\[2mm]
0&0&0&-\textbf{i}\xi^t\\[2mm]
\textbf{i}\xi d_2&0&-\textbf{i}\xi b_2&-I_{3}
\end{array}\right)\]
with
$$b_1=1+|\xi|^2+\frac{2}{|\xi|^2},\ \ d_1=\frac{2}{|\xi|^2},
\ \ b_2=1+|\xi|^2+\frac{1}{|\xi|^2},\ \ d_2=\frac{1}{|\xi|^2},\ \
I_3=\mbox{diag}(1,1,1)
$$ and the notation \textbf{i} is the imaginary unit. Here
$\hat{J}_a=(\hat{J}^{(1)}_a,\hat{J}^{(2)}_a,\hat{J}^{(3)}_a)$,
$\hat{J}_b=(\hat{J}^{(1)}_b,\hat{J}^{(2)}_b,\hat{J}^{(3)}_b)$. \\
We solve the O.D.Es (\ref{4.3}) straightforward by linear O.D.Es
theory and get its solution denoted by
\be
\hat{U}=e^{\hat{A}t}U_0
\label{4.3'}\ee
 where $\hat{U}=(\hat{W}_a,\hat{J}_a,\hat{W}_b,\hat{J}_b)$ with
 \bma
 \hat{W}_a(\xi,t)
=&\frac{1}{6}\hat{W}_{a0}[F_1+2F_2+e_1^-+e_1^++2(e_2^-+e_2^+)]\nnm\\
& +\frac{1}{3}\hat{W}_{b0}[F_1-F_2+e_1^-+e_1^+-(e_2^-+e_2^+)]\nnm\\
&-\frac{\textbf{i}}{3}(\hat{J}_{a0}\cdot\xi)(F_1+2F_2)
-\frac{\textbf{i}}{3}(\hat{J}_{b0}\cdot\xi)(2F_1-2F_2),\label{4.4}
\\
 \hat{W}_b(\xi,t)
=&\frac{1}{6}\hat{W}_{b0}[2F_1+F_2+2(e_1^-+e_1^+)+(e_2^-+e_2^+)]\nnm\\
&+\frac{1}{6}\hat{W}_{a0}[F_1-F_2+e_1^-+e_1^+-(e_2^-+e_2^+)]\nnm\\
&-\frac{\textbf{i}}{3}(\hat{J}_{b0}\cdot\xi)(2F_1+F_2)
-\frac{\textbf{i}}{3}(\hat{J}_{a0}\cdot\xi)(F_1-F_2),\label{4.5}
\ema
and for $k=1,2,3$
\bma  \hat{J}_a^{(k)}(\xi,t)
=
&\frac{\hat{J}_{a0}^{(k)}}{|\xi|^2}(|\xi|^2-\xi_k^2)e^{-t}-\frac{\xi_k}{|\xi|^2}(\sum^3_{\stackrel{l=1}{l\neq
k}}\xi_l\hat{J}_{a0}^{(l)})e^{-t}\nnm\\
&-\frac{\xi_k}{6|\xi|^2}(\xi\cdot\hat{J}_{a0})[2F_2+F_1-2(e_2^-+e_2^+)-(e_1^++e_1^-)]\nnm\\
&+\frac{\xi_k}{3|\xi|^2}(\xi\cdot\hat{J}_{b0})[F_2-F_1+e_1^-+e_1^+-e_2^--e_2^+]\nnm\\
&-2(\hat{W}_{a0}-\hat{W}_{b0})\frac{\textbf{i}\xi_k}{|\xi|^2}F_2-\frac{\textbf{i}}{3}\hat{W}_{a0}\xi_k(1+|\xi|^2)(2F_2+F_1)\nnm\\
&-\frac{2\textbf{i}}{3}\hat{W}_{b0}\xi_k(1+|\xi|^2)(F_1-F_2),\label{4.6}
\\
\hat{J}_b^{(k)}(\xi,t) =
&\frac{\hat{J}_{b0}^{(k)}}{|\xi|^2}(|\xi|^2-\xi_k^2)e^{-t}
-\frac{\xi_k}{|\xi|^2}(\sum^3_{\stackrel{l=1}{l\neq
k}}\xi_l\hat{J}_{b0}^{(l)})e^{-t}\nnm\\
&-\frac{\xi_k}{6|\xi|^2}(\xi\cdot\hat{J}_{b0})[F_2+2F_1-(e_2^-+e_2^+)-2(e_1^++e_1^-)]\nnm\\
&+\frac{\xi_k}{6|\xi|^2}(\xi\cdot\hat{J}_{a0})[F_2-F_1+e_1^-+e_1^+-e_2^--e_2^+]\nnm\\
&+(\hat{W}_{a0}-\hat{W}_{b0})\frac{\textbf{i}\xi_k}{|\xi|^2}F_2
 -\frac{\textbf{i}}{3}\hat{W}_{b0}\xi_k(1+|\xi|^2)(F_2+2F_1)\nnm\\
&-\frac{\textbf{i}}{3}\hat{W}_{a0}\xi_k(1+|\xi|^2)(F_1-F_2)\label{4.7}\ema
where
 \be
e_1^-=e^{-\frac{t}{2}(1-I_1)},~~e_1^+=e^{-\frac{t}{2}(1+I_1)},
~~e_2^-=e^{-\frac{t}{2}(1-I_2)},~~e_2^+=e^{-\frac{t}{2}(1+I_2)}\label{4.8}
\ee with \be
I_1=\sqrt{1-4|\xi|^2(1+|\xi|^2)},~~~I_2=\sqrt{1-4(3+|\xi|^2(1+|\xi|^2))},
\label{4.9}\ee and \be
F_1=\frac{e^{-\frac{t}{2}(1-I_1)}-e^{-\frac{t}{2}(1+I_1)}}{I_1},~~~
~F_2=\frac{e^{-\frac{t}{2}(1-I_2)}-e^{-\frac{t}{2}(1+I_2)}}{I_2}.
\label{4.10} \ee Note here that we have
$E_0=\nabla\triangle^{-1}(W_{a0}-W_{b0})\in L^2(\mathbb{R}^3)$ which
implies
$(\hat{W}_{a0}-\hat{W}_{b0})\frac{\textbf{i}\xi_k}{|\xi|^2}\in
L^2(\mathbb{R}^3)$ in \ef{4.6}--\ef{4.7}. This means the existence
of the inverse transformation of $\hat{U}$ and thus the global
solvability of $U$ for \ef{4.2}--\ef{4.2'}.
\par
\bigskip

\underline{\textit{Proof of Theorem~\ref{theorem 1.3}}}.\quad We
first focus on the estimates of the lower bound in
\ef{1.18'}--\ef{1.19'}. The idea is to analyze the Fourier
transformation of $U$ due to the Plancherel theorem. In view of
\ef{4.8}--\ef{4.10} we should give some properties of the terms
contained in $W_a,J_a,W_b,J_b$ given by \ef{4.4}--\ef{4.7}.
\par

\textbf{We have the following estimates
\begin{eqnarray}
& |e_1^+|+|e_2^-|+|e_2^+|+|F_2|+|\xi |^2|F_2|<ce^{-ct}, &for\quad
\xi\in R^3, \label{4.20}
\\[2mm]
&|e_1^-|\leq e^{-\frac{t}{2}},\quad |F_1|\leq \frac t2e^{-\frac12t},
\quad |\xi|^2|F_1|<c\frac t2e^{-\frac12t}, \quad  & for \quad
|\xi|^2\geq \mbox{$\frac{\sqrt{2}-1}{2}$}, \label{4.21}
\\[2mm]
&e_1^-\geq e^{-c|\xi|^2t}, & for\quad    |\xi|^2\leq
 \mbox{$\frac{\sqrt{2}-1}{2}$}, \label{4.22}
\end{eqnarray}
where and below $c>0$ is a generic positive constant. The estimates
(\ref{4.20}) is gained by a direct computation. The estimates
(\ref{4.21}),(\ref{4.22}) can be obtained as follows}.
\par

It holds for $|\xi|^2\geq \frac{\sqrt{2}-1}{2}$ that
$$
1-4|\xi|^2(1+|\xi|^2)\leq0,~~I_1=\sqrt{1-4|\xi|^2(1+|\xi|^2)}
=\textbf{i}\sqrt{4|\xi|^2(1+|\xi|^2)-1}=\textbf{i}|I_1|$$ and
$$
|e_1^-|=|e^{-\frac{t}{2}(1-I_1)}|\leq e^{-\frac{t}{2}}.
$$
By
$$
|F_1|=|\frac{e^{-\frac{t}{2}(1-I_1)}-e^{-\frac{t}{2}(1+I_1)}}{I_1}|
=|\frac
t2e^{-\frac{t}{2}}(\frac{e^{\textbf{i}\frac{t|I_1|}{2}}-e^{-\textbf{i}\frac{t|I_1|}{2}}}{\textbf{i}\frac{t|I_1|}{2}})|~~~~
\text{and}~~~~|(e^{\textbf{i}s})^{'}| \leq 1,
$$
we know
$$
|F_1|\leq \mbox{$\frac{t}{2}$}e^{-\frac12t}.
$$
As for $|\xi|^2|F_1|$, it holds for
$\frac{\sqrt{2}-1}{2}\leq|\xi|^2<\frac{\sqrt{3}-1}{2}$  that
$$
|\xi|^2|F_1|\leq c\mbox{$\frac{t}{2}$}e^{-\frac12t}.
$$
When $\frac{\sqrt{3}-1}{2}\leq|\xi|^2$, we can directly compute
$$
|\xi|^2|F_1|=|\xi|^2|e^{-\frac{t}{2}}(\frac{e^{\textbf{i}\frac{t|I_1|}{2}}-e^{-\textbf{i}\frac{t|I_1|}{2}}}{\textbf{i}|I_1|})|
=e^{-\frac
t2}|\frac{|\xi|^2}{{\textbf{i}|I_1|}}||(e^{\textbf{i}\frac{t|I_1|}{2}}-e^{-\textbf{i}\frac{t|I_1|}{2}})|\leq
c e^{-\frac t2}.
$$

By the fact that $1-\sqrt{1-4s(1+s)}\leq 2(\sqrt{2}+1)s$ for $0\leq
s\leq \frac{\sqrt{2}-1}2$, we can obtain (\ref{4.22}) easily since
$e_1^-=e^{-\frac{t}{2}(1-\sqrt{1-4|\xi|^2(1+|\xi|^2)})}\geq
e^{-c|\xi|^2t}$.
\par

With the help of \ef{4.20}--\ef{4.22} we can turn to calculate the
time-decay rates of density and momentum $\hat{W}_a$, $\hat{W}_b$,
$\hat{J}_a$, $\hat{J}_b$, and we take $\hat{W}_a,\hat{J}_a$ for
simplicity. Set
$$
\hat{W}_a=T_1+R_1,\quad \hat{J}_a^{(k)}=T_2^{(k)}+R_2^{(k)},\quad
k=1,2,3
$$
with
 \bma
&T_1=\frac16(\hat{W}_{a0}+2\hat{W}_{b0})(F_1+e_1^-)\label{4.23}\\
&R_1=\hat{W}_a-T_1\qquad (\text{the\  rest\  terms})\label{4.24}\\
& T_2^{(k)}
 =-\frac{\textbf{i}}3(\hat{W}_{a0}+2\hat{W}_{b0})\xi_k(1+|\xi|^2)F_1
   \label{4.25}\\
&R_2^{(k)}=\hat{J}_a^{(k)}-T_2^{(k)}\qquad (\text{the\ rest\
terms})\label{4.26}
 \ema
\par

  By
\ef{4.20}--\ef{4.22} we know
 \bma
\|\hat{W}_a(.,t)\|^2&
 \geq \frac12\int_{R^3}|T_1|^2d\xi-\int_{R^3}|R_1|^2d\xi\nnm\\
 &\geq\frac12\int_{|\xi|^2< \frac{\sqrt{2}-1}{2}}|\frac16(\hat{W}_{a0}+2\hat{W}_{b0})(F_1+e_1^-)|^2d\xi-c(1+t)e^{-ct}\nnm\\
 &\geq\frac12\int_{|\xi|^2< \frac{\sqrt{2}-1}{2}}|\frac16(\hat{W}_{a0}+2\hat{W}_{b0})e_1^-|^2d\xi-c(1+t)e^{-ct}\nnm\\
&\geq\int_{|\xi|^2< \min
\{\frac{\sqrt{2}-1}{2},r^2\}}ce^{-2c|\xi|^2t}d\xi
 -c(1+t)e^{-ct}\nnm\\
&\geq c(1+t)^{-\frac32} -c(1+t)e^{-ct}\label{4.27}\ema where we have
used the assumption in Theorem \ref{theorem 1.3} that
$|(\hat{W}_{a0}+2\hat{W}_{b0})|>m_0>0$ in $B(0,r)$ and the fact
 $F_1>0$ for $|\xi|^2<\frac{\sqrt{2}-1}{2}$.
We also used $(|\xi|^n \hat{W}_{a0},|\xi|^n \hat{W}_{b0},|\xi|^l
\hat{J}_{a0},|\xi|^l \hat{J}_{b0})\in L^2(R^3) $ for the integers
$0\leq n\leq6,~~0\leq l\leq5$ and \ef{1.17'} to get
$\int_{R^3}|R_1|^2d\xi<ce^{-ct}$ with the help of (\ref{4.20}). The
above $c>0$ denotes the generic positive constant depending on the
norm of initial data and $m_0$ and not necessarily be the same.\par

The combination of Plancherel theorem and inequality (\ref{4.27})
implies for  $t\gg1$ that
 \be
\|W_a(.,t)\|=\|\hat{W}_a(.,t)\|\geq c_1(1+t)^{-\frac34}
\label{4.28} \ee with $c_1$ some positive number. Similarly, with
the help of (\ref{4.20})-(\ref{4.22}), we have
 \bma
 \|\textbf{i}\xi_k \hat{W}_a(.,t)\|^2
&
\geq\frac12\int_{R^3}|\textbf{i}\xi_kT_1|^2d\xi-\int_{R^3}|\textbf{i}\xi_kR_1|^2d\xi\nnm\\
&\geq
    \frac12\int_{|\xi|^2< \frac{\sqrt{2}-1}{2}}|
     \frac{\xi_k}6(\hat{W}_{a0}+2\hat{W}_{b0})(F_1+e_1^-)|^2d\xi
 -c(1+t)e^{-ct}\nnm\\
&\geq
   c\int_{|\xi|^2<\min \{\frac{\sqrt{2}-1}{2},r^2\}}|\xi_k|^2|e^{-2c|\xi|^2t}|d\xi
  -c(1+t)e^{-ct}\nnm\\
&\geq
  c(1+t)^{-\frac52} -c(1+t)e^{-ct}. \label{4.29}
 \ema
 It follows from (\ref{4.29})
that
 \be
\|\partial_{x_k}W_a(.,t)\|^2=\|\textbf{i}\xi_k\hat{W}_a(.,t)\|^2\geq
c_1(1+t)^{-\frac54}\label{4.30} \ee for $t\gg1$. Repeating the
similar procedure as above, we can estimate the higher order term
$\|\textbf{i}^{|\alpha|}\xi_1^{\alpha_1}\xi_2^{\alpha_2}\xi_3^{\alpha_3}
\hat{W}_a\|^2$ $(|\alpha|\leq6)$, which together with the Plancherel
theorem leads to the algebraic time-decay rate for ${W}_a$ from
below \be
 \|\partial_x^l W_a(.,t)\|_{L^2(\mathbb{R}^3)}
 \geq c(1+t)^{-\frac{l}{2}-\frac34}
 ,\ \ 0\leq l\leq6.  \label{4.31}\ee
\par

Again, we can repeat the similar argument as above to establish the
corresponding algebraic time-decay rate for $\hat{J}_a$. In fact, by
(\ref{4.20})-(\ref{4.22}) we have after a direct computation that
 \bma
\|\hat{J}_a^{(k)}(.,t)\|^2
&\geq\frac12\int_{R^3}|T_2^{(k)}|^2d\xi-\int_{R^3}|R_2^{(k)}|^2d\xi\nnm\\
&\geq\frac12\int_{R^3}|T_2^{(k)}|^2d\xi-c(1+t)e^{-ct}\nnm\\
&\geq c\int_{|\xi|^2<\min
\{\frac{\sqrt{2}-1}{2},r^2\}}|\xi_kF_1|^2d\xi-c(1+t)e^{-ct},
  \qquad\quad  k=1,2,3.\label{4.32}
\ema Note that $\frac{5-2\sqrt{2}}{4}<I_1<1$ for
$0\leq|\xi|^2<\frac12(\frac{\sqrt{2}-1}{2})$, we have
 \be
|F_1|=|\frac{e^{-\frac{t}{2}(1-I_1)}-e^{-\frac{t}{2}(1+I_1)}}{I_1}|
= \frac1{|I_1|}|e^{-\frac t2(1-I_1)}(1-e^{-tI_1})|\geq ce^{-\frac
t2(1-I_1)}    \label{4.32'} \ee for $t>1$ and
$|\xi|^2<\frac12(\frac{\sqrt{2}-1}{2})$. By (\ref{4.22}) and the
fact $e^{-\frac t2(1-I_1)}\geq ce^{-c|\xi|^2t}$ for
$|\xi|^2\leq\frac12(\frac{\sqrt{2}-1}{2})$, we finally obtain from
\ef{4.32'} that \be |F_1|\geq ce^{-c|\xi|^2t},\quad
\text{for}\quad |\xi|^2<\mbox{$\frac12(\frac{\sqrt{2}-1}{2})$}.
\label{4.32''} \ee Set
$r_1^2=\min\{r^2,\frac12(\frac{\sqrt{2}-1}{2})\}$ and let $t>1$.
By (\ref{4.32}), (\ref{4.32''}), we have
 \bma
 \|\hat{J}_a^{(k)}(.,t)\|^2
&\geq
  c\int_{|\xi|^2<r_1^2}|\xi_k|^2e^{-2c|\xi|^2t}d\xi-c(1+t)e^{-ct}\nnm\\
&\geq c(1+t)^{-\frac52}-c(1+t)e^{-ct},\qquad \quad  k=1,2,3.
\label{4.33} \ema This gives rise to the time-decay rate of
$J_a=({J}^{(1)}_a,{J}^{(2)}_a,{J}^{(3)}_a)$ for $t\gg1$ that
 \be
\|J_a(.,t)\|=\|\hat{J}_a(.,t)\| \geq c_1(1+t)^{-\frac54}.
\label{4.34} \ee
\par

The higher order estimates of $J_a$ can be established in the
similar argument as obtaining \ef{4.29} for $W_a$ and finally we can
have for  $t\gg1$ that
 \be
\|D_x^lJ_a(.,t)\|\geq c(1+t)^{-\frac
l2-\frac54},\label{4.35}\qquad \quad \l=1,2,3,4,5.
 \ee
The above estimates are valid for $W_b,J_b^{(k)}(k=1,2,3)$ due to
the symmetry between $W_a$ and $W_b$, $J_a$ and $J_b$. Thus the
proof of the lower bound estimates in Theorem \ref{theorem 1.3} is
finished.
 \par

Note that the time-decay rate  from above of solutions in
(\ref{1.18'}) and (\ref{1.19'}) can be obtained in the same
framework of Fourier transformation to establish the lower bound of
decay rate. Also, it can be obtained by energy methods used in
Section 3, we omit the details.~~ $\square$

~~ \\[3mm]

\noindent\textbf{\textsl{Acknowledgements :}} The authors
acknowledge the partial support by the National Science Foundation
of China (No.10571102), the Key Research Project on Science and
Technology of the Ministry of Education of China (No.104072), the
grant- NNSFC (No.10431060),  Beijing Nova program, and the Re Shi Bu
Ke Ji Ze You program.

\end{document}